\documentclass[aps,eqsecnum,nofootinbib,showpacs]{revtex4}
\usepackage[dvips]{graphicx}  
\usepackage{amsmath,amsfonts}
\usepackage{bm}
\newcommand{\romanNum}[1]{\@roman{#1}}
\usepackage{color}  
\def\be{\begin{eqnarray}}
\def\ee{\end{eqnarray}}
\def\beq{\begin{equation}}
\def\eeq{\end{equation}}

\def\p{\partial}

\def\({\left (}
\def\){\right )}
\def\[{\left [}
\def\[{\right ]}

\newcommand{\IP}[2]{\big\langle\, #1\, \big\vert\, #2\, \big\rangle}
\newcommand{\Exp}[1]{\langle\, #1\, \rangle}
\bmdefine{\bmk}{\bm{k}} \bmdefine{\bmx}{\bm{x}}
\bmdefine{\bmA}{\bm{A}} \bmdefine{\bmB}{\bm{B}}
\bmdefine{\bmJ}{\bm{J}}
\newcommand{\calL}{\mathcal{L}}
\newcommand{\calO}{\mathcal{O}}
\begin{document}

\author{Hua-Bi Zeng$^{1}$, Zhe-Yong Fan$^{1}$, and Hong-Shi Zong$^{1,2}$}
\address{$^{1}$ Department of Physics, Nanjing University, Nanjing 210093, China}
\address{$^{2}$ Joint Center for Particle, Nuclear Physics and Cosmology, Nanjing 210093, China}
\title{Superconducting Coherence Length and Magnetic Penetration Depth of a $p$-wave Holographic Superconductor}

\begin{abstract}
A classical $SU(2)$ Einstein-Yang-Mills theory in 3+1 dimensional
 anti-de Sitter spacetime is believed to be dual to a $p$-wave superconductor
  in 2+1 dimensional flat spacetime. In order to calculate the superconducting
  coherence length $\xi$ of the holographic superconductor near the superconducting
  phase transition point, we study the perturbation of the gravity theory analytically.
  The superconductiong coherence length $\xi$ is found to be proportional to $(1-T/T_c)^{-1/2}$ near the
  critical temperature $T_c$. We also obtain the magnetic penetration depth $\lambda\propto(T_c-T)^{1/2}$ by adding
  a small external homogeneous magnetic field. The results agree with the Ginzburg-Landau theory.
\end{abstract}
 \maketitle

\section{ introduction}

The AdS/CFT correspondence \cite{maldacena,gubser1,witten,aharony}
has played an important role in understanding strongly coupled gauge
theories. Recently, it also has been applied to superconductivity.
The key point of the holographic theories for superconductors is
that in the gravity theory, a black hole coupled with matter fields
will have symmetry breaking solutions. There are mainly two
holographic models of superconductors with different matter sectors.
The first one is an Abelian-Higgs model which is the gravity dual of
an $s$-wave superconductor with a scalar order parameter. The
properties of this holographic superconductor model have been
studied by many authors [5-31]. The other one is an
Einstein-Yang-Mills (EYM) theory in which the condensate carries
angular momenta [32-43].

Gubser \cite{gubser4} firstly presented an argument that by coupling
the Abelian Higgs model to gravity with a negative cosmological
constant, one can get solutions which spontaneously break the
Abelian gauge symmetry via a charged complex scalar condensate near
the horizon of the black hole. Hartnoll {\it et al} \cite{hartnoll1}
explored this Abelian-Higgs model of superconductivity further. They
built an $s$-wave holographic (in the sense of AdS/CFT duality)
superconductor with scalar order parameter which exhibits the basic
features of a superconductor such as the existing of a critical
temperature below which a charged condensate forms. The behaviors of
the $s$-wave holographic superconductor under magnetic field have
been studied in many papers
\cite{nakano,albash1,maeda1,hartnoll2,Albash2,Albash3,Montull,Maeda3}.
Especially, Maeda and Okamura \cite{maeda1} studied the
superconducting coherence length $\xi$ of the $s$-wave holographic
superconductor near the critical temperature $T_c$. They obtained
that $\xi$ is proportional to $(1-T/T_c)^{-1/2}$, which is in
agreement with the Ginzburg-Landau theory.

The other holographic superconductor model which is an EYM model
with fewer parameters whose Lagrangian is determined by symmetry
principles is constructed later by Gubser \cite{gubser5} and is
shown to have spontaneous symmetry breaking solutions due to a
condensate of non-Abelian gauge fields in the theory. Gubser and
Pufu studied this model with both $p$-wave backgrounds and
$(p+ip)$-wave backgrounds \cite{gubser6}. Roberts and Hartnoll
studied the $(p+ip)$-wave backgrounds and found two major
nonconventional features for this holographic superconductor that
are different from the $s$-wave counterpart. One is the existence of
a pseudogap at zero temperature, and the other is the spontaneous
breaking of time reversal symmetry \cite{roberts}. In our recent
paper \cite{Zeng}, we studied the phase transition properties of
this model in constant external magnetic field. We found that the
added magnetic field indeed suppresses the superconductivity. In the
present paper, we further study the $p$-wave holographic
superconductor composed of a non-Abelian gauge fields (the matter
sector) and a black hole background (the gravity sector) by using
perturbation theory near the critical temperature, following closely
Maeda and Okamura \cite{maeda1}. According to the Ginzburg-Landau
theory, the superconducting length, or the correlation length of the
order parameter is an important characteristic parameter for a
superconductor. Since the order parameter of the $p$-wave
holographic superconductor is the vector operator dual to the
charged non-Abelian gauge field, we investigate the static
fluctuation of the condensed non-Abelian gauge field with
nonvanishing spatial momentum along one spatial direction of the AdS
boundary to get the correlation length $\xi$. A homogenous magnetic
field in the field theory is added by placing a small vector
potential in the matter sector. The magnetic penetration length
$\lambda$ is obtained by calculating the London current of the
holographic superconductor.

The organization of this paper is as follows. In Section II, we
reconstruct the superconducting solution of the EYM theory which is
dual to a $p$-wave superconductor by perturbation techniques.
Section III is devoted to the derivation of $\xi$ by solving the
eigenvalue equations from the perturbation. In Section IV we find
that the London current can be induced by a homogeneous magnetic
field, and the magnetic penetration length is also studied. The
conclusion and some discussions are given in Section V.

\section{Model of a $p$-wave holographic superconductor}
In this section, we review the gravity dual theory of the $p$-wave
superconductor. The starting point of studying holographic
superconductor at finite temperature $T$ is choosing a black hole
solution with a negative cosmological constant such that the Hawking
temperature of the black hole is $T$. The full EYM theory in 3+1
dimensional spacetime considered in Refs.
\cite{gubser5,gubser6,roberts} has the following action
\begin{eqnarray}
S_{\textmd{EYM}}=
\int\sqrt{-g}d^4x\left[\frac{1}{2\kappa_4^2}\left(R+\frac{6}{L^2}\right)
-\frac{L^2}{2g_{\rm YM}^2}\textmd{Tr}(F_{\mu\nu}F^{\mu\nu})\right],
\end{eqnarray}
where $g_{\rm YM}$ is the gauge coupling constant and
$F_{\mu\nu}=T^aF^a_{\mu\nu}=\partial_\mu A_\nu-\partial_\nu
A_\mu-i[A_\mu,A_\nu]$ is the field strength of the gauge field
$A=A_\mu dx^\mu=T^aA^a_\mu dx^\mu$. For the $SU(2)$ gauge symmetry,
$[T^a,T^b]=i\epsilon^{abc}T^c$ and
$\textmd{Tr}(T^aT^b)=\delta^{ab}/2$, where $\epsilon^{abc}$ is the
totally antisymmetric tensor with $\epsilon^{123}=1$. The Yang-Mills
Lagrangian becomes
$\textmd{Tr}(F_{\mu\nu}F^{\mu\nu})=F^a_{\mu\nu}F^{a\mu\nu}/2$ with
the field strength components $F^a_{\mu\nu}=\partial_\mu
A^a_\nu-\partial_\nu A^a_\mu+\epsilon^{abc}A^b_\mu A^c_\nu$.

Working in the probe limit in which the matter fields do not
backreact on the metric as in Refs. \cite{gubser5,gubser6,roberts}
and taking the planar Schwarzchild-AdS  ansatz, the  black hole
metric reads (we use mostly plus signature for the metric)
\begin{equation}
ds^2=-f(r)dt^2+\frac{dr^2}{f(r)}+\frac{r^2}{L^2}(dx^2+dy^2),
\label{metric}
\end{equation}
where the metric function $f(r)$ is
\begin{equation}
f(r)=\frac{r^2}{L^2}(1-\frac{r_0^3}{r^3}).
\end{equation}
$L$ and $r_0$ are the radius of the AdS spacetime and the horizon
radius of the black hole, respectively. They determine the Hawking
temperature of the black hole,
\begin{equation}
T=\frac{3r_0}{4\pi L^{2}},
\end{equation}
which is also the temperature of the dual gauge theory living on the
boundary of the AdS spacetime.
 Now we introduce a new coordinate $z=r_0/r$. The metric (\ref{metric}) then becomes
 \begin{equation}
ds^2=\frac{L^2\beta^2(T)}{z^2}(-h(z)dt^2+dx^2+dy^2)+\frac{L^2dz^2}{z^2h(z)},
 \end{equation}
where $h(z)=1-z^3$ and $\beta(T)=r_0/L^2=4\pi T/3$.

Using the Euler-Lagrange equations, one can obtain the equations of
motion for the gauge fields,
\begin{equation}
\frac{1}{\sqrt{-g}}\partial_{\mu}\left(\sqrt{-g}F^{a\mu\nu}\right)
+\epsilon^{abc}A^{b}_{u}F^{c\mu\nu}=0.
\end{equation}
For the $p$-wave backgrounds, the ansatz \cite{gubser6} takes the
following form,
\begin{equation}
A=\phi(z)T^3dt+w(z)T^1dx.
\end{equation}
With this ansatz, we can derive the equations of motion for the two
dimensionless quantities $\tilde{w}(z)=w(z)/\beta(T)$ and
$\tilde{\phi}(z)=\phi(z)/\beta(T)$,
\begin{equation}
\frac{d}{dz}(h(z)\frac{d\tilde{w}}{dz})+\frac{\tilde{\phi}^2\tilde{w}}{h(z)}=0,
\label{eom2}
\end{equation}
and
\begin{equation}
\frac{d^2\tilde{\phi}}{dz^2}-\frac{\tilde{\phi}\tilde{w}^2}{h(z)}=0.
\label{eom1}
\end{equation}
Here the $U(1)$ subgroup of $SU(2)$ generated by $T^3$ is identified
to the electromagnetic gauge group \cite{gubser5} and $\phi$ is
the electrostatic potential, which must vanish at the horizon for
the gauge field $A$ to be well-defined, but need not vanish at
infinity. Thus the black hole can carry charge through the
condensate $w$, which spontaneously breaks the $U(1)$ gauge
symmetry. This is a Higgs mechanism, but there are Goldstone bosons
corresponding to changing the directions of the condensate in real
space or gauge space. They must be visible in the bulk as normal
modes or (more likely) quasi-normal modes.

 The exact solution of the equations of motion is
\begin{equation}
\tilde{w}=0, \tilde{\phi}=\mu/\beta(T)-qz=q(1-z),
\end{equation}
where $\mu$ is interpreted as the chemical potential of the field
theory. This trivial solution is parameterized by the dimensionless
constant $q$, which is related to the charge density of the dual
field theory which couples to $\mu$ as
\begin{equation}
\langle J^{0}\rangle=\frac{\delta S_{\text{on-shell
boundary}}}{\delta A_0^3}=\frac{1}{2g_{\rm YM}^2}\beta^2(T)q.
\end{equation}

The superconducting solution with non-vanishing $\tilde{w}$ takes
the following asymptotic form at the AdS boundary,
\begin{align}
  & \tilde{w}= \frac{\langle
\calO\rangle}{\sqrt{2}}z
  + \cdots,
\end{align}
\begin{align}
  & \tilde{\phi} = \mu/\beta(T) - q z + \cdots,
\end{align}
where $\langle \calO\rangle$ is the condensate  of the charged
operator dual to the field $w$ and is the order parameter for the
superconductivity phase. We demand that the constant term vanish
since we require that there is no source term in field theory action
for the operator $\langle \calO \rangle$ \cite{gubser6,roberts}.

According to numerical calculations \cite{gubser6}, the order
parameter behaves as
\begin{equation}
\langle\calO\rangle \sim(1-T/T_c)^{1/2}
\end{equation}
near the critical phase transition point. For the reason of
continuity, the solution at the critical temperature should be
\begin{equation}
\tilde{w}_c=0, \tilde{\phi}_c=q_c(1-z).
\end{equation}
The non-trivial solution near the critical temperature can be
obtained by a perturbation expansion in terms of
$\epsilon=(1-T/T_c)$ since $\epsilon$ is a small parameter. We
expand $\tilde{w(z)}$ and $\tilde{\phi(z)}$ as
\begin{equation}
   \tilde{w}(z)
  = \epsilon^{1/2}\, \tilde{w}_1(z)
  + \epsilon^{3/2}\, \tilde{w}_2(z) + \cdots~,
  \label{exp1}
\end{equation}
\begin{align}
 \tilde{\phi}(z)
  = \tilde{\phi}_c(z) + \epsilon\, \tilde{\phi}_1(z) + \cdots~.
\label{exp2}
\end{align}

Substituting Eq.(\ref{exp1}) and Eq.(\ref{exp2}) into
Eq.(\ref{eom1}) and Eq.(\ref{eom2}), we obtain equations for
$\tilde{w}_1$ and $\tilde{\phi}_1$,
\begin{equation}
{\calL}_w\tilde{w}_1(z)=0,
\label{eq.expand1}
\end{equation}
\begin{equation}
\frac{d^2\tilde{\phi}_1(z)}{dz^2}-\frac{\tilde{\phi}_c(z)\tilde{w}_1^2(z)}{h(z)}=0,
\label{eq.expand2}
\end{equation}
where we have defined the following operator,
\begin{equation}
{\cal
L}_w=-\left(\frac{d}{dz}h(z)\frac{d}{dz}+\frac{\tilde{\phi}_c^2(z)}{h(z)}\right).
\end{equation}

\section{The Superconducting Coherence Length }
 As an important parameter for superconductor, the superconducting
 coherence length is obtained from the complex pole of the static
 correlation function of the order parameter in Fourier space:
\begin{align}
  & \Exp{\Tilde{\calO}(\Vec{k}\,) \Tilde{\calO}(-\Vec{k}\,)}
  \sim \frac{1}{\vert\, \Vec{k}\, \vert^2 + 1/\xi^2}~.
\end{align}
The pole $\vert\, \Vec{k}\, \vert^2$ can be calculated in the probe
limit by perturbing the fields ($\tilde{w}$, $\tilde{\phi}$) in the
equations of motion Eq.(\ref{eom1}) and Eq.(\ref{eom2}). It is
enough to consider a linear perturbation with fluctuation of the
field in the $y$-direction which takes the following form,
\begin{equation}
\delta \tilde{\phi}(z,y)dt=[\Phi(z,k)dt]e^{iky},
\end{equation}
\begin{equation}
\delta \tilde{w}(z,y)=[W(z,k)]e^{iky}.
\end{equation}

Using this perturbation, we get the following linearized equations
for $W$ and $\Phi$:
\begin{equation}
\tilde{k}^2W=(\frac{d}{dz}h(z)\frac{d
}{dz}+\frac{\tilde{\phi}^2}{h(z)})W+\frac{2\tilde{w}\tilde{\phi}}{h(z)}\Phi,
\end{equation}

\begin{equation}
\tilde{k}^2\Phi=(h(z)\frac{d^2}{dz^2}-\tilde{w}^2)\Phi-2\tilde{\phi}\tilde{w}W,
\end{equation}
where $\tilde{k}=k/\beta(T)$ is dimensionless.

Now, our task is to solve the eigenvalue equations near $T_c$
analytically. Using the perturbation expansions in Eq.(\ref{exp1})
and Eq.(\ref{exp2}), we get
\begin{equation}
-\tilde{k}^2W=({\calL}_w-\frac{2\epsilon\tilde{\phi}_c\tilde{w}_1}{h(z)})W-\frac{2\epsilon^{1/2}\tilde{\phi}_c\tilde{w}_1}{h(z)}\Phi,
\label{eq.eigein1}
\end{equation}

\begin{equation}
-\tilde{k}^2\Phi=(-h(z)\frac{d^2}{dz^2}+\epsilon\tilde{w}_1^2)\Phi+2\epsilon^{1/2}\tilde{\phi}_c\tilde{w}_1W.
\label{eq.eigein2}
\end{equation}
The boundary conditions for the two equations are
\begin{equation}
W(1)=\textrm{regular}, ~~~~~\Phi(1)=0
\end{equation}
at the horizon and
\begin{equation}
W(z)=(\textrm{const})\times z+O(z^2)~ ,
\end{equation}
\begin{equation}
\Phi(z)=(\textrm{const})\times z+O(z^2)~
\end{equation}
near the AdS boundary $z=0$.

 The trivial solution is the zeroth
order solution $\Phi_0$ and $W_0$ with $\tilde{k}=0$,
\begin{equation}
\Phi=0, ~~~~W_0=\tilde{w}_1,
\end{equation}
where equation (\ref{eq.expand1}) is used. The non-trivial solutions
can be obtained by a series expansion around the zeroth order
solution in $\epsilon$,
\begin{equation}
W=\tilde{w}_1+\epsilon W_1+\epsilon^2 W_2+\cdots,
\end{equation}
\begin{equation}
\Phi=\epsilon^{1/2}\Phi_1+\epsilon^{3/2}\Phi_2+\cdots,
\end{equation}
\begin{equation}
 \tilde{k}^2=\epsilon\tilde{k}^2_1+\epsilon^2\tilde{k}^2_2+\cdots.
\end{equation}
Using this expansion in Eq.(\ref{eq.eigein1}) and
Eq.(\ref{eq.eigein2}), one has
\begin{equation}
-\tilde{k}^2_1
\tilde{w}_1={\calL}_wW_1-\frac{2\tilde{\phi_c}\tilde{w}_1}{h(z)}(\tilde{\phi}_1+\Phi_1),
\label{eq.k}
\end{equation}
\begin{equation}
\frac{d^2\Phi_1}{dz^2}=\frac{2\tilde{\phi}_c\tilde{w}_1^2}{h(z)}=\frac{2d^2\tilde{\phi}_1}{dz^2}.
\end{equation}
Eq. (\ref{eq.k}) can be solved for $\tilde{k}$ by defining an inner
product for the states $w_I$ and $w_{II}$,
\begin{equation}
 \IP{w_I}{w_{II}}= \int_0^1 dzw_I^*(z)~w_{II}(z).
\end{equation}
Using this inner product for Eq. (\ref{eq.k}) and $\tilde{w_1}$,
with the fact that ${\cal L}_w\tilde{w}_1=0$, we have
\begin{equation}
\tilde{k}^2_1\IP{\tilde{w}_1}{\tilde{w}_1}=\left\langle
\tilde{w}_{1}|\frac{2\tilde \phi_{c}\tilde{w}_{1}}{h(z)} \tilde
\phi_{1}\right\rangle +2\int_{0}^{1}dz\frac{\tilde
\phi_{c}\tilde{w}_{1}^{2}}{h(z)}\Phi_{1}.
\end{equation}
The first term of the above equation vanishes, which can be seen from
the Hermiticity of ${\cal L}_w$ and
\begin{equation}
{\cal L}_w\tilde{w}_{2}=\frac{2\tilde
\phi_{c}\tilde{w}_{1}}{h(z)}\tilde \phi_{1}. \label{eq.w2}
\end{equation}
Eq.(\ref{eq.w2}) is the equation of motion for $\tilde{w}_2$ defined
in (\ref{exp1}). Using the fact that $\tilde k^{2}=\epsilon \tilde
k_{1}^{2}$, the eigenvalue $\tilde k$  in a  first order
approximation can be written as
 \begin{equation}
 \tilde k^{2}=\epsilon\frac{N}{D} + O(\epsilon^{2}),
 \end{equation}
 where
 \begin{eqnarray}
 N=2\int_{0}^{1}dz\frac{\tilde \phi_{c}\tilde{w}_{1}^{2}}{h(z)}\Phi_{1}~~~ \textmd{and} ~~~D=\int_{0}^{1}dz\psi_{1}^{2}.
 \end{eqnarray}

Finally, the superconducting coherence length is given by
\begin{equation}
\xi = \frac{\epsilon^{-1/2}}{\beta(T_c)}\sqrt{\frac{D}{N}}
+O(\epsilon^{2})\propto \left( 1 - \frac{T}{T_c} \right)^{-1/2}~.
\end{equation}
We have thus obtained the same critical exponent $(-1/2)$ for $\xi$ as given by the
Ginzburg-Landau theory.

\section{The London Equation And Magnetic Penetration Length}
In order to calculate the magnetic penetration length for the
holographic superconductor, we add a homogenous external magnetic
field by assuming a perturbative non-zero $\delta
A_{y}^{3}(z,x)=b(z)x$, where $ \lim_{z \to 0}\, \delta
A_{y}^{3}(z,x)=Bx$. Then the magnetic field in the field theory is
$F_{xy} = \p_x \delta A_{y}= B$ \cite{maeda1}. We still work in the
probe limit where the magnetic field does not affect the metric. If
we only focus on the neighborhood of $x=0$, the equation of motion
for $b(z)$ can be treated as decoupled from $\tilde{w}$,
\begin{equation}
(\frac{d}{dz}h(z)\frac{d}{dz}-\tilde{w}^2)b(z)=0,
 \label{eq.b}
\end{equation}
where $b(z)$ must satisfie the regularity boundary condition at the
horizon $z=1$. This equation can also be solved by perturbation. We
can expand $b(z)$ as
\begin{equation}
b(z)=b_0(z)+\epsilon b_1(z)+ \cdots.
\end{equation}
Substituting this expansion and Eq.(\ref{exp1}) into
Eq.(\ref{eq.b}), we obtain the equations,
\begin{equation}
\frac{d}{dz}h(z)\frac{d}{dz}b_0(z)=0, \label{eq.b1}
\end{equation}

\begin{equation}
\frac{d}{dz}h(z)\frac{d}{dz}b_1(z)-\tilde{w}_1^2(z)b_0(z)=0.
\label{eq.b2}
\end{equation}
The solution of Eq.(\ref{eq.b1}), which satisfies the
required boundary conditions is

\begin{equation}
b_0(z)=C,
\end{equation}
where $C=B$ is a constant since the condition $\lim_{z \to
0}\,b(z)=B$ must be satisfied. So the solution of Eq.(\ref{eq.b2})
should be,

\begin{equation}
\frac{d b_1}{dz}=-\frac{B}{h(z)}\int_z^1dz_0\tilde{w}_1^2(z_0).
\end{equation}
Integrating the above equation. we have,
\begin{equation}
b(z)=B-\epsilon B\int^z_0 \frac{dz_1}{h(z_1)}
    \int^1_{z_1} dz_0~\tilde{w}_1^2(z_0)
  + O(\epsilon^2).
  \label{eq.b(z)}
\end{equation}
Using the fact that $B = \lim_{z \to 0}\, b(z)$ and $\delta
A_y^{3(0)}(x) = \lim_{z \to 0}\, \delta A_y^3(z, x)$, we can rewrite
Eq.(\ref{eq.b(z)}) as
\begin{equation}
\delta A_y^3(z, x) =
          \delta A_z^{3(0)}(x)\,\left( 1
  - \, \epsilon\, \int^z_0 \frac{dz_1}{h(z_1)}
    \int^1_{z_1} dz_0~\tilde{w}_1^2(z_0) \right)
  + O(\epsilon^2).
\end{equation}

According to the AdS/CFT dictionary, we can read out the current
$\langle J_y(x)\rangle$ near $T_c$ to be,
\begin{flalign}
\langle J_y(x)\rangle=-\frac{L^2}{g_{YM}^2}(\frac{4\pi
T_c}{3})(1-\frac{T}{T_c})\int_0^1dz{\tilde w}_1^2(z)\delta
A_y^{3(0)}(x) + O(\epsilon^2),
\end{flalign}
or
\begin{equation}
\langle J_y(x)\rangle \sim-T_c\epsilon\delta A_y^{3(0)}(x).
\label{eq.J}
\end{equation}
This is similar to the London equation for real world
superconductors,
\begin{align}
  & \bmJ = - \frac{e_*^2}{m_*}~\psi^2\, \bmA
  = - e_*\, n_s\, \bmA~,
\label{eq:London_eq}
\end{align}
where $e_*$ and $m_*$ are effective charge and mass of the order
parameter respectively, and $n_s$ is the superfluid number density.

Comparing Eq.(\ref{eq.J}) and Eq.(\ref{eq:London_eq}), we find
that the superfluid density $n_s$ near the critical point in the
field theory is
\begin{align}
  & n_s \sim \epsilon\, T_c \sim T_c-T~.
\label{magnetic penetration depth}
\end{align}
According to the Ginzburg-Landau theory, the magnetic penetration
depth $\lambda$ is given by
\begin{align}
  & \lambda \sim 1/\sqrt{n_s}~.
\end{align}
Then, we get the behavior of $\lambda$ in the vicinity of the
critical temperature,
\begin{equation}
\lambda\propto(T_c-T)^{-1/2},
\end{equation}
which is the expected result as in the Ginzburg-Landau theory.

\section{ conclusion and discussions}
For the EYM theory with a $p$-wave backgrounds, we investigated the
linear fluctuation of the condensation solution under the probe
limit. By solving the linearized eigenvalue equations with only
spatial momentum along one spatial direction by the perturbation
method, we obtain that the correlation length $\xi$ diverges as
$\xi\sim(1-T_c/T)^{-1/2}$ at the critical temperature. We also find
that the magnetic penetration length behaves as
$\lambda\sim(T_c-T)^{-1/2}$ near the critical temperature. The
London type equation Eq.(\ref{eq.J}) implies a Meissner effect in
the superconductor. These results are consistent with the
Ginzburg-Landau theory, which supports the idea that the non-Abelian
holographic model can be used to describe superconductors. Our
results are similar to those of the $s$-wave holographic
superconductor studied by Maeda and Okamura \cite{maeda1}. Though
the holographic models have made many achievements, it is still a
unsolved question why the Ginzburg-Landau behavior is
expected and when one would expect deviations from it in these
models. Recently, vortex solutions of the $s$-wave holographic
superconductor in homogeneous external magnetic field have been
studied in Ref. [16-19]. It will be interesting to study possible
localized vortex solutions in the $p$-wave holographic
superconductor.

\section{acknowledgement}

We thank professor C. P. Herzog and professor S. S. Gubser for help.
We also thank Wei-Ming Sun for helpful discussions. This work is
supported in part by the National Natural Science Foundation of
China (under Grant No. 10775069) and the Research Fund for the
Doctoral Program of Higher Education (Grant Nos. 20060284020 and
20080284020).


\begin{thebibliography}{}\label{sec:TeXbooks}


\bibitem{maldacena}
J.~M.~Maldacena, ``The large N limit of superconformal field
theories and supergravity,'' Adv.\ Theor.\ Math.\ Phys.\  {\bf 2}
(1998) 231 [arXiv:hep-th/9711200].
\bibitem{gubser1}
S.~S.~Gubser, I.~R.~Klebanov and A.~M.~Polyakov,
 ``Gauge Theory Correlators from Non-Critical String Theory,''
 Phys.\ Lett.\ B {\bf428}, 105 (1998)
 [arXiv:hep-th/9802109].
\bibitem{witten}
E.~Witten,
 ``Anti-de Sitter space and holography,
  '' Adv.\ Theor.\ Math.\ Phys.\  {\bf 2}, 253 (1998)
   [arXiv:hep-th/9802150].
\bibitem{aharony}
O.~Aharony, S.~S.~Gubser, J.~M.~Maldacena,
  H.~Ooguri, and Y.~Oz,
 ``Large N Field Theories, String Theory and Gravity,''
  Phys.\ Rept.\ {\bf323}, 183 (2000)
 [arXiv:hep-th/9905111].




\bibitem{hartnoll1}
 S.~A.~Hartnoll, C.~P.~Herzog and G.~T.~Horowitz,
  ``Building a Holographic Superconductor,''
  Phys.\ Rev.\ Lett.\  {\bf 101}, 031601 (2008)
  [arXiv:0803.3295 [hep-th]].




\bibitem{nakano}
E.~Nakano and W.~Y.~Wen,
  ``Critical Magnetic Field In A Holographic Superconductor,''
  Phys.\ Rev.\  D {\bf 78}, 046004 (2008)
  [arXiv:0804.3180 [hep-th]].

\bibitem{albash1}
T.~Albash and C.~V.~Johnson,
  ``A Holographic Superconductor in an External Magnetic Field,''
  JHEP {\bf 0809}, 121 (2008)
  [arXiv:0804.3466 [hep-th]].

\bibitem{maeda1}
 K.~Maeda and T.~Okamura,
  ``Characteristic length of an AdS/CFT superconductor,''
  Phys.\ Rev.\ D {\bf 78}, 106006 (2008)
  [arXiv:0809.3079 [hep-th]].

\bibitem{hartnoll2}
 S.~A.~Hartnoll, C.~P.~Herzog and G.~T.~Horowitz,
  ``Holographic Superconductors,''
  JHEP {\bf 0812}, 015 (2008)
  [arXiv:0810.1563 [hep-th]].

\bibitem{herzog1}
C.~P.~Herzog, P.~K.~Kovtun and D.~T.~Son,
  ``Holographic model of superfluidity,''
  Phys.\ Rev.\ D {\bf79}, 066002 (2009)
  [arXiv:0809.4870 [hep-th]].

\bibitem{Horowitz1}
G.T. Horowitz and M.M. Roberts,
  \lq\lq Holographic Superconductors with Various Condensates,\rq\rq
  Phys.\ Rev.\ D {\bf78} 126008 (2008) [arXiv:0810.1077 [hep-th]].

\bibitem{Franco}
S.~Franco, A.~Garcia-Garcia, D.~Rodriguez-Gomez ,``A general class
of holographic superconductors,'' [arXiv:0906.1214 [hep-th]].

\bibitem{Amado}
I.~Amado, M.~Kaminski, K.~Landsteiner, ``Hydrodynamics of
Holographic Superconductors,'' JHEP {\bf 0905}, 021 (2009).

\bibitem{Maeda2}
K.~Maeda, M.~Natsuume, T.~Okamura, ``Universality class of
holographic superconductors,'' Phys.\ Rev.\ D {\bf79}, 126004
(2009).

\bibitem{Kim}
Youngman Kim, Yumi Ko, Sang-Jin Sin,`` Density driven symmetry
breaking and Butterfly effect in holographic superconductors ,''
[arXiv:0904.4567 [hep-th]].

\bibitem{Albash2}
 T.~Albash, C.~V.~Johnson,
 ``Phases of Holographic Superconductors in an External
 Magnetic,''[arXiv:0906.0519 [hep-th]].

\bibitem{Albash3}
 T.~Albash, C.~V.~Johnson,
 ''Vortex and Droplet Engineering in Holographic Superconductors,'' [arXiv:0906.1795 [hep-th]].

\bibitem{Montull}
M.~Montull, A.~Pomarol, P.~J.~Silva,``The Holographic Superconductor
Vortex,'' Phys.\ Rev.\ Lett {\bf103} 091601 (2009) [arXiv:0906.2396
[hep-th]].

\bibitem{Maeda3}
K.~Maeda, M.~Natsuume, T.~Okamura,'' ``Vortex lattice for a
holographic superconductor,'' [arXiv:0910.4475 [hep-th]].

\bibitem{Keranen3}
V.~Keranen, E.~Keski-Vakkuri, S.~Nowling, K. P.
Yogendran,``Inhomogeneous Structures in Holographic Superfluids: II.
Vortices,'' [arXiv:0912.4280 [hep-th]].


\bibitem{Nishioka}
T.~Nishioka, S.~Ryu, T.~Takayanagi,``Holographic
Superconductor/Insulator Transition at Zero Temperature,''
[arXiv:0911.0962 [hep-th]].

\bibitem{Konoplya}
A. Konoplya, A. Zhidenko,``Holographic conductivity of zero
temperature superconductors,'' [arXiv:0909.2138 [hep-th]].

\bibitem{Horowiitz2}
  G.~T.~Horowitz, M.~M.~
Roberts,`` Zero Temperature Limit of Holographic Superconductors,''
JHEP 11 (2009) 015 [arXiv:0908.3677  [hep-th]].

\bibitem{Gubser2}
S.~S.~Gubser, A.~Nellore,``Ground states of holographic
superconductors,'' [arXiv:0908.1972 [hep-th]].

\bibitem{Sin}
 S.~J.~Sin, S.~S.~Xu, Y.~Zhou,`` Holographic Superconductor for a Lifshitz fixed
point,'' [arXiv:0909.4857 [hep-th]].

\bibitem{Brynjolfsson}
 E.J.Brynjolfsson, U.H. Danielsson, L. Thorlacius, T. Zingg,``
Holographic Superconductors with Lifshitz Scaling,''[arXiv:0908.2611
[hep-th]].

\bibitem{Umeh}
O.~C.~Umeh,``Scanning the Parameter Space of Holographic
Superconductors,'' JHEP 0908: 062,2009 [arXiv:0907.3136 [hep-th]].

\bibitem{Herzog2}  C.~P.~Herzog, A.~Yarom,``Sound modes in holographic
superfluids,''[arXiv:0906.4810 [hep-th]].

\bibitem{Keranen1}
V.~Keranen, E.~Keski-Vakkuri, S.~Nowling, K. P. Yogendran,``Dark
Solitons in Holographic Superfluids,'' [arXiv:0906.5217v3 [hep-th]].

\bibitem{Keranen2}
V.~Keranen, E.~Keski-Vakkuri, S.~Nowling, K. P.
Yogendran,``Inhomogeneous Structures in Holographic Superfluids: I.
Dark Solitons,'' [arXiv:0911.1866 [hep-th]].

\bibitem{Chen}
J. W. Chen, Y. J. Kao and W. Y. Wen,``Peak-Dip-Hump from Holographic
Superconductivity,'' [arXiv:0911.2821 [hep-th]].


%
\bibitem{gubser6}
 S.~S.~Gubser and S.~S.~Pufu,
  ``The gravity dual of a p-wave superconductor,''
  JHEP {\bf 0811}, 033 (2008)
  [arXiv:0805.2960 [hep-th]].

\bibitem{roberts}
 M.~M.~Roberts and S.~A.~Hartnoll,
  ``Pseudogap and time reversal breaking in a holographic superconductor,''
  JHEP {\bf 0808}, 035 (2008)
  [arXiv:0805.3898 [hep-th]].

\bibitem{ammon1}
 M.~Ammon, J.~Erdmenger, M.~Kaminski and P.~Kerner,
 ``Superconductivity from gauge/gravity duality with flavor,'' Phys. Lett. {\bf B680} (2009)
 [arXiv:0810.2316 [hep-th]].

\bibitem{ammon2}
 M.~Ammon, J.~Erdmenger, M.~Kaminski and P.~Kerner,
 ``Flavor Superconductivity from Gauge/Gravity Duality,'' JHEP 10 (2009) 067
 [arXiv:0903.1864 [hep-th]].

\bibitem{basu2}
 P.~Basu, J.~He, A.~Mukherjee and H.~H.~Shieh,
 ``Superconductivity from D3/D7: Holographic Pion Superfluid,''
 [arXiv:0810.3970 [hep-th]].

\bibitem{herzog3}
 C.~P.~Herzog and S.~S.~Pufu,
  ``The Second Sound of SU(2),'' JHEP 04 (2009) 126
  [arXiv:0902.0409 [hep-th]].



\bibitem{sonner}
 J.~Sonner,
 ``A Rotating Holographic Superconductor,''Phys.\ Rev.\ D {\bf80}, 084031
  [arXiv:0903.0627 [hep-th]].

\bibitem{Zeng}
H.~Zeng, Z.~Fan, Z.~Ren, ``Time Reversal Symmetry Breaking
Holographic Superconductor in Constant External Magnetic
Field,''Phys.\ Rev.\ D {\bf80}, 066001 [arXiv:0903.2323 [hep-th]].

\bibitem{Peeters}
K.~Peeters, J.~Powell, M.~Zamaklar,`` Exploring colourful
holographic superconductors,'' JHEP {\bf0909}, 101, 2009
[arXiv:0907.1508 [hep-th]].

\bibitem{Herzog4}
C.~P.~Herzog,``Lectures on Holographic Superfluidity and
Superconductivity,''J. Phys. A {\bf42} (2009) 343001
[arXiv:0904.1975 [hep-th]].

\bibitem{Basu}
Pallab.~Basu, Jianyang.~He, Anindya.~Mukherjee, Hsien-Hang.~Shieh,
`` Hard-gapped Holographic Superconductors,'' arXiv:0911.4999.

\bibitem{Ammon}
Martin Ammon, Johanna Erdmenger, Viviane Grass, Patrick Kerner, Andy
O'Bannon,`` On Holographic p-wave Superfluids with Back-reaction,''
arXiv:0912.3515.

\bibitem{gubser4}
S.~S.~Gubser,
 ``Breaking an Abelian gauge symmetry near a black
   hole horizon,''
   Phys.\ Rev.\  D {\bf 78}, 065034 (2008)
  [arXiv:0801.2977 [hep-th]].
\bibitem{gubser5}
 S.~S.~Gubser,
  ``Colorful horizons with charge in anti-de Sitter space,''
  Phys.\ Rev.\ Lett.\  {\bf 101}, 191601 (2008)
  [arXiv:0803.3483 [hep-th]].
%



\end{thebibliography}
\end{document}